# Short Secret Sharing Using Repeatable Random Sequence Generators

Arvind Srinivasan

SplitByte Inc., Los Gatos, CA 95030, USA

Chien-Chung Chan

Department of Computer Science, The University of Akron, Akron, OH 44325, USA

*Abstract*— We present a new secret sharing algorithm that provides the storage efficiency of an Information Dispersal Algorithm (IDA) while providing perfect secret sharing. We achieve this by mixing the input message with random bytes generated using Repeatable Random Sequence Generator (RRSG). We also use the data from the RRSG to provide random polynomial evaluation points and optionally compute the polynomials on random isomorphic fields rather than a single fixed field.

## I. INTRODUCTION

With the availability of highly scalable public object stores and file storage to enterprises and developers, there is a move to store data in a highly distributed manner. Ability to store and access data from anywhere to anywhere else in the world does provide new security challenges [1]. We address the creation and retrieval of data in large scale by describing a new storage efficient secret sharing algorithm with no special needs to store encryption key externally.

Secret sharing has become very important in solving many of today's data storage and computation problems. It is an approach by which data can be split into $n$ shares and can be recovered by having any of $m$ ($\leq n$) shares. Since Shamir [2] originally published his secret sharing algorithm, several research articles have appeared in extending the algorithm and finding new applications. A general summary of these developments can be found in Biemel [3]. In this paper, we will focus our attention to the ($n$, $m$) threshold use case, although the algorithm can be easily extended to general access use cases.

The original Shamir's algorithm used polynomials of order $m$-1 as in the following Equation (1) to perform secret sharing. The secret is encoded in the constant coefficient $a_0$, the remaining coefficients are selected randomly and the resulting polynomial is evaluated at $n$ distinct points to produce $n$ shares.

$$f(x) = \sum_{i=0}^{m-1} a_i x^i . \qquad (1)$$

Typically, the computations are performed over a finite field, more specifically Galois Field of order $2^b$ represented as $GF(2^b)$. Given a message of size $s$ bytes, this method requires $n*s$ bytes to store the secret. However the method of Rabin's Information Dispersal Algorithm (IDA) [4] can store the same information using $(n*s)/m$ bytes. This storage savings come at the expense of reduced security. Blakley & Meadow [5] described a ramp approach that traded-off security and storage, and this has come to be known as the generalized Shamir Secret Sharing or Packed Secret Sharing. Basically, the secrets are embedded in ($a_0$, $a_1$, .. $a_r$) and all other coefficients ($a_{r+1},...a_{m-1}$) are chosen randomly. Note when $r=(m-1)$, the ramp method becomes a pure IDA scheme and $r=0$ would yield the traditional Shamir secret sharing. The storage used with a ramp scheme is $(n*s)/(r+1)$ bytes. Krawczyk [6] combined symmetric encryption and secret sharing to provide the security of secret sharing and storage cost of IDA. In this method, a random encryption key is generated and is used to encrypt the secret. The random encryption key is split into $n$ shares using the traditional Shamir's method and the encrypted data is split using traditional IDA method. Since the key length is minimal compared to actual secret, the overhead is very minimal. Further, theoretical bounds of each share size for various access structure can be found in [7], [8].

There is another method in [9] that used a specific IDA methodology namely, error correcting Reed Solomon erasure coding to produce a threshold scheme. In order to get security, they used All or Nothing Encryption of Rivest [10] instead of secret sharing. The resulting algorithm has provided sizes of shares similar to that of [6] while giving security guarantees.

In this paper, we propose a secret sharing method that would provide the storage efficiency of the IDA algorithm while providing the security of Shamir's secret sharing. We achieve this by using random values for point of evaluations *x* and evaluating the polynomials using a random isomorphic field instead of a fixed one.

## II. PRELIMINARIES

One of the first needs is to make some assumptions on the secret data entropy. For example, we do not want all data bytes to be zero. Our approach is to mix the secret data with a random byte stream. Actually, one can see the encryption step in Krawczyk [6] as a preprocessing step. With sufficient encryption strength, the entropy of input data is increased. The second idea is to make evaluation points of each share to be a random value. The only requirement of Shamir's is that the *x* values have to be unique, i.e., we need *n* unique *x* values. The randomness in *x* will make the algorithm to be secret sharing rather than an IDA algorithm. More importantly, the randomness of *x* increases the security even in traditional Shamir's method as opposed to the standard use of fixed *x* values. The third idea is to make use of random isomorphic field over which the polynomials are computed. We will abstract the random value generation problem in the next Section and then we will describe the algorithm.

## III. REPEATABLE RANDOM SEQUENCE GENERATOR (RRSG)

We define the term Repeatable Random Sequence Generator (RRSG) as a random byte sequence capable of being repeated by knowing a set of initial states or keys. More formally, we define RRSG *R* as a random sequence of *t* bytes denoted by $\{r_0, r_1, r_2, ..., r_{t-1}\}$ and can be reproduced again given an initial state or key. In terms of entropy, the same can be stated as

$H(R) \sim$ *close to maximum H of uniform distribution*
$H(R \mid key) = 0$ (2)

RRSG can be seen as an abstraction of many concepts we have previously seen in the literature. These include:
- All strong PRNG [11],[12] automatically are RRSG as the initial seed (key) once known renders the sequence known.
- Stream Ciphers such as Salsa and ChaCha family [13], FCSR [14] are RRSG with the same initial states.
- All strong Symmetric Ciphers such as AES [15] that encrypts any well known data sequence (e.g. all bytes are 0x05) are RRSG with the initial state being the (Encryption Key, Initialization Vector or IV).
- All Mask Generating Function as defined in [16].
- Mixing two or more RRSG also produces an RRSG.

RRSG can be randomly seekable if we can generate any $r_i$, $\forall\ i > 0$ by starting the generation from $r_j$ with $(i\text{-}j) < M$ where *M* is a small positive integer constant. This property will allow us to reconstruct any random subsection of data without having to regenerate RRSG bytes from the beginning.

## IV. GALOIS FIELD

Galois fields are finite fields represented by $GF(p^n)$, where *p* is any prime number and n is any integer greater than or equal to 1. In particular, $GF(2^8)$ is used in many applications including the popular symmetric encryption method, AES. The fields in the class $GF(2^n)$ are considered in this section. Addition and its inverse (subtraction) are defined as the bitwise XOR operator and multiplication is defined as the polynomial multiplication divided by a prime polynomial. In general, there are more than one such polynomial resulting in a rich set of isomorphic fields. The number of irreducible polynomials of order n is given by the following Gauss Formula [17],

$$\frac{1}{n} \sum_{d \mid n} \mu(n/d) p^{nd} \qquad (3)$$

where μ is the Mobius function and *d* denotes the number of all divisors of *n* including 1 and *n*. The table below shows the count of prime polynomial for various order of Galois Field. As expected, the number of prime polynomial grows rapidly as the GF order increases.

TABLE 1. Number of prime polynomials in different order.

| $GF(2^n)$ | Number Primes | $GF(2^n)$ | Number Primes | $GF(2^n)$ | Number Primes |
|---|---|---|---|---|---|
| $GF(2^4)$ | 3 | $GF(2^{13})$ | 630 | $GF(2^{22})$ | 190,557 |
| $GF(2^5)$ | 6 | $GF(2^{14})$ | 1,161 | $GF(2^{23})$ | 364,722 |
| $GF(2^6)$ | 9 | $GF(2^{15})$ | 2,182 | $GF(2^{24})$ | 698,870 |
| $GF(2^7)$ | 18 | $GF(2^{16})$ | 4,080 | $GF(2^{25})$ | 1,342,176 |
| $GF(2^8)$ | 30 | $GF(2^{17})$ | 7,710 | $GF(2^{26})$ | 2,580,795 |
| $GF(2^9)$ | 56 | $GF(2^{18})$ | 14,532 | $GF(2^{27})$ | 4,971,008 |
| $GF(2^{10})$ | 99 | $GF(2^{19})$ | 27,594 | $GF(2^{28})$ | 9,586,395 |
| $GF(2^{11})$ | 186 | $GF(2^{20})$ | 52,377 | $GF(2^{29})$ | 18,512,790 |
| $GF(2^{12})$ | 335 | $GF(2^{21})$ | 99,858 | $GF(2^{30})$ | 35,790,267 |

## V. ALGORITHM

Let the secret data of $N$ words be represented as $\{d_0, d_1, d_2, ..., d_{N-1}\}$ with each word having a size of $b$ bits with all calculations done in one or more isomorphic fields of $GF(2^b)$. Our goal is to create a threshold scheme such that the data is split into $n$ parts and requiring at least $m$ parts to reconstruct. We choose a RRSG method (e.g. Salsa20). We generate the keys and other initial states such as Initialization Vector(IV) randomly and collectively call them as Key $K$. The exact size of this information is dependent on the choice of RRSG method. The various steps of the storing process are as follows:

(a) We use traditional Shamir [2] to split $K$ into $n$ secret shares as $\{k_0, k_1, k_2, ..., k_{n-1}\}$. Note we can use any isomorphic field of $GF(2^b)$, and each $k_i$ is equal to the size of the Key $K$.

Next we repeat steps (b) to (e) for every $m$ words. If the data size is not a multiple of $m$, suitable padding method is used. Set each share $S_i = \{k_i\}$

(b) For every $m$ words, we generate ($m+n+4$) random words $R$ as $\{r_0, r_1, r_2, ..., r_{m+n+3}\}$ using RRSG.

(c) Next, we set the polynomial coefficients as $a_i = r_i \oplus d_i \ \forall \ i \in \{0, m-1\}$, where $\oplus$ is the XOR operator.

(d) Let $X = \{x_0, x_1 ..., x_{n-1}\}$ represent the $n$ distinct points of polynomial evaluation. We set $x_i = 1 + \left(r_{m+i} \bmod \left(2^b - 1\right)\right) \ \forall \ i \in \{0, n-1\}$. We avoid the value $x_i=0$. Also, it is possible that the $x_i$ can repeat and we use the additional 4 random words to make the evaluation point distinct. By using a fixed number of random values, we will be able to randomly seek for a block of byte ranges.

(e) We generate an integer value $I$ using last 4 words $\{r_{m+n}, r_{m+n+1}, r_{m+n+2}, r_{m+n+3}\}$. We use this integer to pick the $GF(2^b)$ isomorphic field. For example, if we use $GF(2^8)$, there are 30 isomorphic fields and we can number each field from 0 to 29. We simply take ($I$ mod 30) as the field of choice.

(f) Now, we evaluate Shamir polynomial represented by equation (1) for all elements of $X$ using the isomorphic field of $GF(2^b)$ picked in step (e) to calculate transformed value $Y = \{y_0, y_1 ..., y_{n-1}\}$.

(g) Update each share $S_i = \{S_i \ || \ y_i\} \ \forall \ i \in \{0, n-1\}$, where $||$ is the concatenation operator.

To reconstruct the data, we need $m$ distinct Shares $S_{t_i}$. Let $T$ be the set of retrieved shares and be denoted by distinct members $\{t_0, t_1, t_2, ..., t_{m-1}\}$ with $t_i \in \{0, 1, 2, ..(n-1)\}$. We then reconstruct the data by applying the following steps:

(a) We recover the set $\{k_{t_0}, k_{t_1}, ..k_{t_{m-1}}\}$ from the shares by reading the first size($K$) bytes from each share. We recover the key $K$ using standard Shamir's approach. We use this to initialize the RRSG function, and now it will be able to produce the same random bytes as in the data split process. Also set the Recovered data as $D = \{\}$.

We repeat the following steps (b) to (f) for every available set of $m$ words after the first $K$ bytes. For the last set, if we detect padding, we will remove the extra padding bytes.

(b) Reading a word from each share, we construct set $Y^R = \{y_{t_0}, y_{t_1} ..., y_{t_{m-1}}\}$. Note each of y's are read from each share available to us for reconstructing the original data.

(c) We use the RRSG in step (a) to generate ($m+n+4$) random words $R$ as $\{r_0, r_1, r_2, ..., r_{m+n+3}\}$.

(d) Next we generate the $n$ distinct points of polynomial evaluation $X = \{x_0, x_1 ..., x_{n-1}\}$. As in the encryption step, we generate the elements as $x_i = 1 + \left(r_{m+i} \bmod \left(2^b - 1\right)\right) \ \forall \ i \in \{0, n-1\}$. In the event the generated elements are not unique, we do the same method of conflict resolution using additional 4 random words to make the evaluation point distinct. We use $X$'s elements to construct $X^R = \{x_{t_0}, x_{t_1} ..., x_{t_{m-1}}\}$ and we determine the isomorphic field id by using the 4 random words as in the encryption step.

(e) We can now use $X^R$ and $Y^R$ to recover the polynomial coefficients $a_i$ using interpolation on the specific isomorphic field.

(f) We recover the data $d_i = r_i \oplus a_i \ \forall \ i \in \{0, m-1\}$ and update the data $D$ by concatenating the recovered data as $D = \{D \ || \ d_i\}$.

## VI. ALGORITHM ANALYSIS

The size of each share is **(n\*s)/m +n\*size(K)**. This is very similar to the size obtained by Krawczyk [6]. We have achieved this while retaining secret-sharing security levels as in [2].

**Theorem 1**: The algorithm described in the previous section is a perfect secret sharing scheme if $p < m$ shares are known.

First the Key **K** is stored using the standard Shamir Secret Sharing [2]. Therefore, it enjoys Perfect Secret Sharing [2], [6], [18]. In other words finding **K** requires at least **m** shares to be available and when this is not the case **K** is completely unknown.

We first treat the case when the choice of RRSG is restricted to strong stream ciphers such as Salsa, ChaCha, or the AES symmetric encryption of well known text. Therefore, when the data is mixed with stream ciphers, the data is encrypted. The second step of polynomial evaluation and splitting can be viewed as an Information Dispersal Algorithm(IDA) of the encrypted data. Therefore, it matches the conditions of [6], and hence the algorithm is secure. Quoting [6] "Knowing (**m**-1) shares does not give any information about other Shares $S_i$ as it does not know about the encrypted stream **E**." In our case the encrypted stream **E** are the coefficients of the polynomial.

We can also prove the same by taking the entropy assumption defined by equation (2) on the RRSG where the unconditional entropy approaching the maximum entropy of uniform distribution, i.e.:

$H(R) \sim$ *close to the maximum H of uniform distribution*

Since we chose the polynomial coefficient $a_i = r_i \oplus d_i$, this implies that the coefficient $a_i$ is also uniformly distributed over all possible values. The evaluation points $X$ is also chosen from the same RRSG and hence $X$ is also uniformly distributed over all possible values. Since both the polynomial and evaluation points are randomly distributed, $Y$ is also uniformly distributed and is true for any isomorphic field. Since we do not know **K**, knowing (**m**-1) share will give us knowledge of $\{y_{t_0}, y_{t_1}, ..., y_{t_{m-2}}\}$, however we do not know the corresponding evaluation points $\{x_{t_0}, x_{t_1}, ..., x_{t_{m-2}}\}$ and the polynomial coefficient are unknown and more specifically, they are both drawn from a uniform distribution. Therefore, knowing y's does not give us any idea of what the coefficients are. This implies the probability distribution $P(a_i | \{y_{t_0}, y_{t_1}, ..., y_{t_{m-2}}\}) = P(a_i)$ which is uniformly distributed and hence we have perfect security. ∎

We now look at a specific variation of algorithm to gain some insight. Instead of using a single RRSG defined by initial state **K**, we use two RRSG with states $(K_1, K_2)$. We use the RRSG characterized by state $K_1$ for encoding the data and the other RRSG using state $K_2$ for generating the evaluation points and the specific isomorphic field. In other words we generate **R** defined as $\{r_0, r_1, r_2, ..., r_{m+n+3}\}$ in step (b) of storing procedure and step (c) of recovery procedure using two different RRSG. Thus $R = (R_1, R_2)$ with $R_1 = \{r_0, r_1, r_2, ..., r_{m-1}\}$ and $R_2 = \{r_m, r_{m+1}, r_{m+2}, ..., r_{m+n+3}\}$ drawn from two different RRSG with initial state $K_1$ and $K_2$ respectively.

**Theorem 2**: The algorithm is a perfect secret sharing scheme if the key $K_1$ is known and up to $p < m$ shares are available to read.

In this case, even though the initial state $K_1$ is known, the resulting polynomial coefficient $a_i = r_i \oplus d_i$ is still more or less uniformly distributed over all values. We still do not know the evaluation points, and even in the worst scenario when (**m**-1) shares are known and hence this give us knowledge of $\{y_{t_0}, y_{t_1}, ..., y_{t_{m-2}}\}$. Therefore following similar arguments in Theorem 1, we have $P(a_i | \{y_{t_0}, y_{t_1}, ..., y_{t_{m-2}}\}) = P(a_i)$ and hence we have perfect secret sharing. ∎

**Theorem 3**: The algorithm is a secure scheme if the key $K_2$ is known and up to (**m**-1) shares are available to read.

This gives us knowledge of $\{x_{t_0}, x_{t_1}, ..., x_{t_{m-2}}, y_{t_0}, y_{t_1}, ..., y_{t_{m-2}}\}$ and the field specification. It is now possible for us to determine certain subset of coefficient $a_i$, however the data $d_i$ itself is unknown as $r_i$ is unknown. Having a strong RRSG and sufficient long key $K_1$, recovering $d_i$ from the coefficient is as difficult as breaking the RRSG. ∎

Using Theorem 2 and 3, if both $K_1$ and $K_2$ are known, then it is possible to recover some of data although not the complete data.

## VII. ISOMORPHIC FIELD

We have a stateful way of determining the specific isomorphic field for computing the shares. Readers should note that this step itself is not needed for the perfect security proof and therefore, is an optional element of the algorithm. It does add a new security feature that works well against the differential cryptanalysis attacks [19]. The stateful or random choice of isomorphic field and the evaluation point *x* will take away advantages an attacker gains by using differential cryptanalysis over brute force attacks. This is particularly true where greater security or diffusion is obtained by doing several rounds of set of operations as in AES. The second advantage is that use of random isomorphic fields does increase the search space. We get 30 fold increase while using $GF(2^8)$ and the same is increased to 4080 fold increase with $GF(2^{16})$. While the search space does increase with higher order of GF, the complexity and running time of algorithm do increase.

## VIII. PERFORMANCE NOTE

The proposed algorithm has similar computation requirements as a standard Shamir's method, however, since we perform packed Shamir, our computation reduces by a factor of ***m***. A good comparison of implementation of various secret sharing algorithms can be seen in [20]. Like many Shamir algorithm implementations that substitute the slow true Random number generation with Pseudo Random numbers, we too benefit by using Stream Cipher and PRNG. Since the current algorithm randomly chooses its evaluation points, the isomorphic fields for computation, we cannot benefit from the computational advantages that one gets by caching Vandermonde matrix. However, this additional computation burden is rewarded with increased security. A similar case can be made for using multiple Isomorphic fields over a single static field. It is worthwhile to note that the $GF(2^8)$ does not add any additional computation load while increasing the search space. However higher orders such as $GF(2^{16})$ increases memory needs for storing discrete logarithm and exponent lookup tables for every isomorphic field that we compute on.
Our initial work primarily focussed on the $GF(2^8)$ field and future work will look at other GF fields.

## IX. CONCLUSION

In this paper, we presented an algorithm that provides true short secret sharing. We achieved this by introducing a RRSG byte stream that is mixed with message data. This increases the entropy of the input message close to the maximum entropy of uniform distribution. We also used the RRSG to provide with random bytes that is used for setting polynomial evaluation points. We introduced the concept of statefully choosing the isomorphic field for computation, and this increases the security of the secret sharing.

It is interesting to note that simply combining the data with random values and doing fully packed Shamir secret sharing actually increased the security level even beyond the original Shamir algorithm. This is due to the fact that all coefficients are random as compared to the original Shamir's method where the constant coefficient is chosen from data. Also, randomly choosing the evaluation points and the isomorphic fields of evaluation give us comparable privacy as the original Shamir's method even when the random values are known. We also leveraged the rich security value that various isomorphic fields provide, and we intuitively suggested that choosing the field in a pseudo-random manner would provide a good defense against differential cryptanalysis attacks. We also believe similar benefit of isomorphic fields can be realized in other areas of cryptography.